# Transition dynamics and metastable states during premelting and freezing of ice surfaces

Shifan Cui*[a], Haoxiang Chen[b]

The premelting of ice is well known, but little is known about how the premelted and solid surfaces convert to each other. In this work, the transition dynamics between two phases are revealed with large-scale molecular dynamics simulations. Supercooling and superheating states exist in the transition, and are overcome by nucleation-like processes. The natural inhomogeneity of ice surfaces enhances nuclei formation, while it only accelerates premelting but not freezing. Furthermore, the complete freezing of ice surfaces may be hindered by the stacking order mismatch between nuclei. This work points out the importance of metastable states in premelting, and the necessity of a large system scale in describing its transition process.

[a]International Center for Quantum Materials, School of Physics, Peking University, 209 Chengfu Road, Haidian District, Beijing 100871, China.
[b]School of Physics, Peking University, 209 Chengfu Road, Haidian District, Beijing 100871, China.
*academic_sfc@outlook.com

## Introduction

The premelting of ice, namely the existence of a quasi-liquid layer (QLL) on the ice surface below the melting point, has been recognized for more than 170 years since Faraday's experiment in 1850 [1]. Due to its important role in various environmental processes [2-5] and the rich physics inside [6-8], premelting has been of research interest for decades [8]. There are considerable studies about the onset of premelting [9-15], the thickness [16-21] and structure [21-27] of QLL, how it behaves during growth or evaporation [28-32], and its interaction with external solvents [33-38] or confinements [39-43].

Despite the progress, there are still several questions about premelting that remain unsolved. There are large discrepancies within the premelting onsets and QLL thicknesses reported in the literature [8, 44], and it is doubted if a particular "transition temperature" exists [45]. Droplets are found on QLL at certain ranges of temperature and vapor pressure [30], but the microscopic mechanism of how they emerge and disappear is still unclear. In addition, the premelting is suggested to explain the well-known slipperiness of ice [46-49], but the scope and validity of such explanations are under challenge [50-52].

Notably, these issues are somewhat related to a less-emphasized topic: how the QLL and unpremelted "solid" surface transition into each other. Indeed, the transition kinetics could be important for interpreting results of various studies about premelting, since metastable phases may play a role in them (just like the bulk solid-liquid transition [53]). Additionally, it may also be relevant to more specific phenomena: the appearance of droplets is related to the ice surface converting between multiple structures [30], and the tribology of ice may be affected by the additional melting induced by sliding and pressures [49, 54-55]. Therefore, understanding how QLL appears and freezes would be helpful for both studying premelting and understanding physics related to premelting. However, few researches have focused on the transition dynamics between QLL and solid surfaces, and none have provided a microscopic description as far as we know.

Molecular dynamics (MD) simulations are widely adapted in premelting research [8], because of their molecule-level resolution and ability to fine-control external conditions. However, the existing MD studies on premelting are usually not able to capture the transition dynamics, due to their limited spatial and temporal scale. In this paper, we present a picture of the transition dynamics between QLL and solid surfaces by large-scale MD simulations. In particular, we observed the supercooling and superheating states during the transition, as an analog to their bulk version. Both directions of transition roughly follow the nucleation picture, but also show qualitatively different behavior from common scenarios. Notably, the surface inhomogeneity of ice induces the formation of certain nuclei, but they only promote the transition from solid surfaces to QLL and not inversely. Furthermore, the mismatch of stacking orders between nuclei can hinder the complete freezing of QLL.

## Methods

Conducting an MD simulation starts with choosing a proper force field. Unluckily, this is not an easy task for studying the premelting of ice, especially when the transition process is of interest. As we will show in this work, the premelting transition involves structural features as large as tens of nanometers, and the process can last for hundreds of nanoseconds. Such large spatial and temporal scaling renders full-atomic models like TIP4P or SPC/E computationally unfeasible. On the other hand, the premelting process involves both bulk and surface molecules and both liquid and solid phases, so the force field needs to be able to describe multiple properties of water and ice. Finally, due to the large discrepancy in experimental results about the premelting of ice [44], their guidance on choosing force fields is limited.

Because of all the reasons above, the optimal choice for this work would be a coarse-grained water model that can reproduce a wide range of physical properties. Coarse-grained water models, by abstracting a water molecule into a single particle, drastically reduce computational cost thus making the large-scale simulations possible. They break the atomic correspondence to the real system, which generally lead to a limited accuracy but can be improved by carefully choosing the function form and parameters. In this work we adopted the ML-mW model [56], a reparameterization of the coarse-grained mW model [57] with improved accuracy on multiple properties. The ML-mW model exhibits a completely premelted first bilayer near the melting point, agreeing with most of experimental results [44] and also the full atomic TIP4P/Ice model (see Appendix A). Its melting point is found to be 295.83±0.5K by the direct coexistence technique [58-59] (Supplemental Material [60], Sec. 1).

To identify how much of the surface is melted, we use the proportion of solid molecules in each bilayer as the order parameter. Namely, each molecule is categorized into one of three types depending on its local structure: (1) hexagonal, (2) cubic, or (3) liquid, by an interleaved version of common neighbor analysis [72]. The order parameter of a bilayer is defined as $1 - N_3/N$, where $N$ is the total number of molecules in the bilayer, and $N_3$ is that in category (3). Therefore, an order parameter of 1.0 indicates a solid ice crystal and 0.0 means the complete loss of order. Compared to some other choices, this method does not need a somewhat arbitrary number to separate liquid and solid states, and is more robust at solid-liquid interfaces. Appendix B presents a detailed description of how this order parameter works, and compares it with several other options.

The majority of MD simulations are performed with LAMMPS [73] and accelerated by INTEL [74] or KOKKOS [75] package, with structural visualizations carried out by OVITO [76]. The TIP4P/Ice [64] simulations for comparison purposes are performed with GROMACS [77]. The detailed parameters of simulations are available in Supplemental Material [60]. In this work we primarily focus on the basal (0001) plane, which is commonly exposed on ice surfaces [78]. Supplemental Material Sec. 9 [60] briefly discusses the premelting on two prismatic ((10-10) and (11-20)) planes.

## Results
### Static premelting profile

Before discussing transition dynamics, it is necessary to determine the static premelting profile: the stable structure of ice surface at given temperatures. To avoid potential interference from metastable states, this is done by an analog to the direct coexistence technique (see Methods section in Supplemental Material [60]).

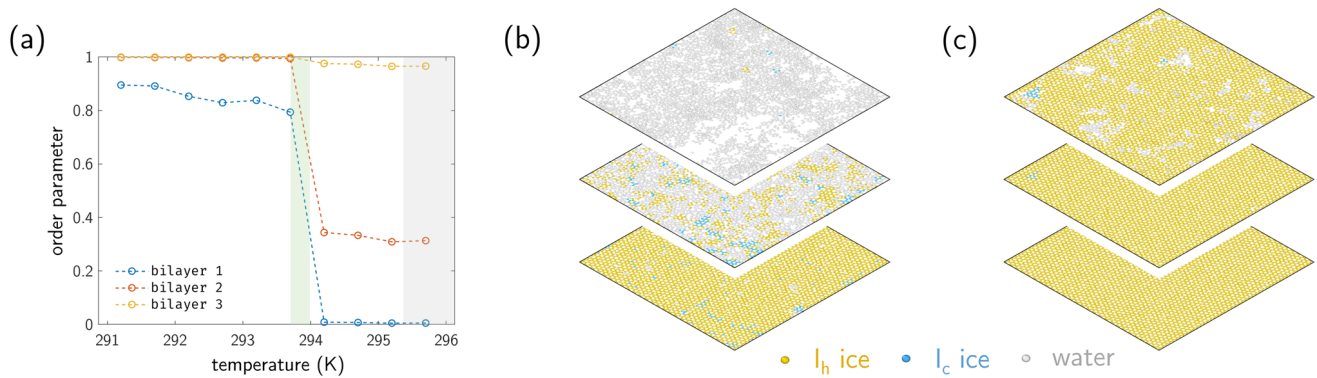

*Figure 1. Overview of the static premelting behavior. (a) the order parameter (proportion of solid molecules) as a function of temperature, for first 3 bilayers on the surface. Gray shadow: estimated melting point of ice. Green shadow: estimated onset of premelting ($T_p$, see Supplemental Material Sec. 2 [60]). (b) the structure of QLL at 294.5K, sliced for first 3 bilayers, in a 20\*20 nm simulation cell. Colors represent the structural type of each water molecule (see legends). (c) the structure of solid surface at 293.1K.*

The results for the ML-mW model are summarized in Fig. 1. There is a single specific onset of premelting temperature for the basal plane, where the first 2 bilayers lose their order simultaneously (Fig. 1(a)). This onset is further determined as $T_p$=293.85±0.15K (Supplemental Material Sec. 2 [60]). The premelting behaves like a 1st-order phase transition, where the order parameters change abruptly at $T_p$. In the premelted surface with T>$T_p$, the 1st bilayer is mostly melted and the 2nd bilayer is mixed with solid and liquid (Fig. 1(b)). In the solid surface with T<$T_p$, only a small fraction of liquid exists in the 1st bilayer and barely any exist in the 2nd bilayer (Fig. 1(c)). The proportion of liquid molecules in the 1st bilayer grows slowly with temperature, but the solid part always dominates when T<$T_p$ (Fig. 1(a)). The premelting mainly involves first 2 bilayers and the 3rd bilayer is mostly intact in both scenarios.

## Metastable states

The 1st-order nature of premelting indicates that metastable, or supercooling/superheating states may exist. To verify this, a series of simulations are performed starting from either solid or premelted surface, in a range of temperatures near $T_p$ (293.1~294.5K). Thermodynamically, the melting of solid surfaces should occur under temperatures higher than $T_p$, and similarly in the opposite direction. Transitions between two phases are indicated by the change of order parameter in first 2 bilayers.

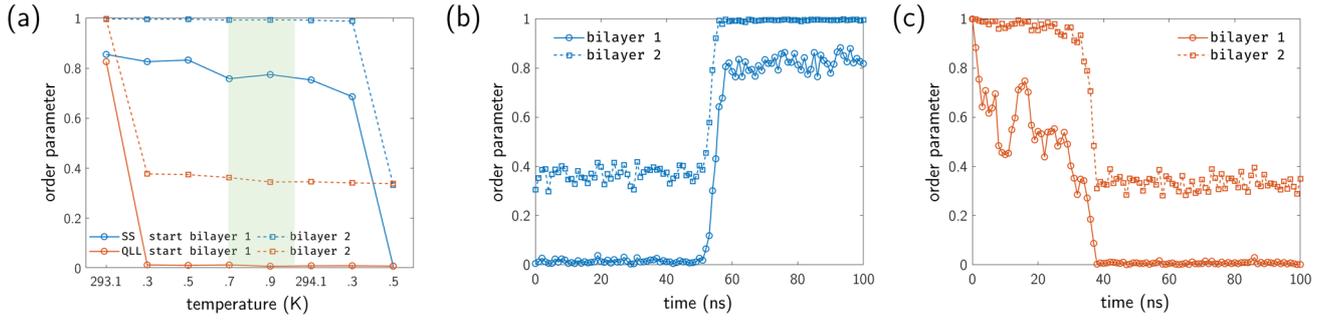

*Figure 2. Supercooling and superheating of the ice surface. (a) the order parameters of first 2 bilayers, at the end of simulations in 20*20 nm systems lasting 100 ns each. QLL start = simulations started from QLL (premelted surface); SS = solid surface. Green shadow: estimated onset of premelting. (b) the order parameters of first 2 bilayers during a QLL -> SS transition at 293.1 K; (c) an SS -> QLL transition at 294.5 K.*

The results of all such trials are summarized in Fig. 2(a). Notably, the transition between two phases does not occur immediately when it is thermodynamically favored. Indeed, the solid surface is stable in a 100 ns simulation up to 294.3K, ~0.5K higher than $T_p$; while the melted surface is stable in 100 ns down to 293.3K, ~0.5K lower than $T_p$ (Fig. 2(a)). These results indicate the existence of supercooling and superheating for the premelting transition, at least for a short time. These metastable phases bring in total uncertainty of ~1K on the premelting onset, which may be relevant to observations about different measurements of QLL during heating and cooling [12, 18]. This uncertainty is smaller than the discrepancy between the $T_p$ reported by various experiments [44], though. Further uncertainty may come from slow dynamics of the transition at lower temperatures, as suggested in another water model [64] (Supplemental Material Sec. 3 [60]).

Fig. 2(b) and (c) demonstrate two instances of transition processes between two phases. Both of them go through some time before an abrupt change of order parameters, confirming the existence of metastable phases. However, the premelting and freezing transitions also show different features: the 1st bilayer of the supercooled phase has a stable order parameter before transition (Fig. 2(b)), but that of the superheated phase fluctuates rather heavily (Fig. 2(c)). This is similar to the impression of its bulk counterpart, where supercooling is considered much more stable than superheating [79]. Such contrast also implies a difference in the transition mechanism between two directions, as we will discuss soon.

## Transition dynamics: premelting

To further investigate the transition mechanism between two phases, we performed large-scale (70*70 nm) MD simulations and visualized the structural change of ice surfaces. The conditions are selected so that (1) the metastable phases can exist for a reasonably long time so they are meaningful; (2) transitions are still observable in the timescale of an MD simulation. Five independent runs are performed in each direction, with different random seeds for initial velocities.

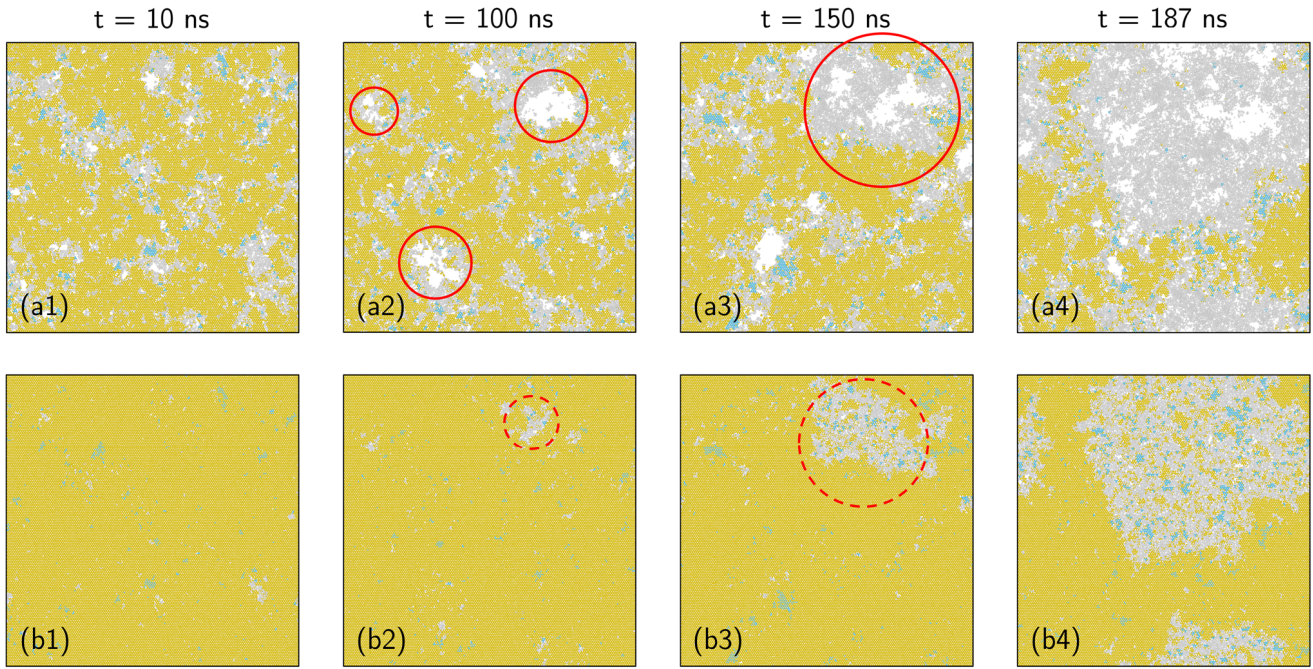

*Figure 3. The structure of ice surface during a solid -> premelted transition. Colors represent different structural types (see legends in Fig. 1). Pure white means no molecules in that region of the bilayer, i.e. holes. The system size is ~70*70 nm. The simulation is performed at 294.2 K, starting from a perfect ice surface. (a1~a4) the 1st bilayer, at 10/100/150/187 ns of the simulation. (b1~b4) the 2nd bilayer. Some holes and liquid nuclei are circled (see main text).*

Fig. 3 shows a typical trajectory of the premelting direction: from solid surface to QLL. In the beginning, the 1st bilayer is a regular mixture of solid and liquid (a1), similar to the equilibrium state at lower temperatures (Fig. 1(c)). After some time, however, holes start to appear in the 1st bilayer (a2, circled); meanwhile, some liquid regions in the 2nd bilayer become larger (b2, circled). One liquid region in the 2nd bilayer grows further later (b3, circled), with larger holes and liquid region in the corresponding 1st bilayer (a3, circled). The liquid region continues to grow (a4, b4) until it occupies the whole surface, completing the phase transition. All five independent runs show a similar procedure of premelting, only with some difference in the transition time (100~250 ns).

The role of holes in premelting deserves attention. Generally speaking, holes are intrinsic features of the ice surface near melting point: they exist both below and above the premelting onset (Fig. 4). Furthermore, they play a special role in premelting: note that the growing liquid region in Fig. 3(b) is located close to a large hole in Fig. 3(a), and it is not a coincidence. Namely, liquid molecules in the 2nd bilayer are more likely to appear near or under holes in the 1st bilayer (Supplemental Material Sec. 4 [60]). In other words, holes lead to a surface inhomogeneity favoring the premelting transition, which requires the 2nd bilayer to lose most of its order. This effect arguably comes from two aspects: (1) holes in the 1st bilayer directly expose those regions of the 2nd bilayer to the air, making them more vulnerable to thermal perturbations; (2) liquid regions are sometimes found near holes in the 1st bilayer (see Fig. 3(a2) and 4(a), for example), so the region beneath them more likely consists of liquid too.

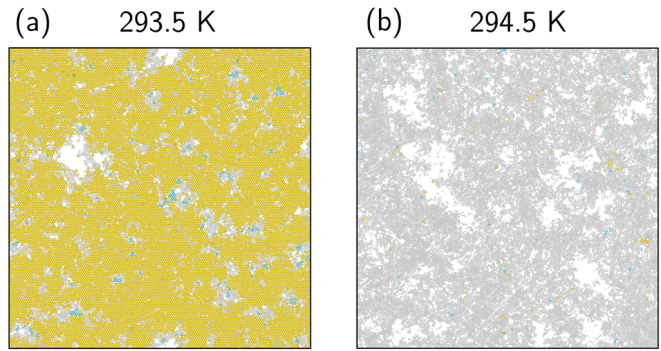

*Figure 4. Structures of the 1st bilayer, in ~70*70 nm simulation cells: (a) the solid surface at 293.5K; (b) QLL at 294.5K. Note that both structures are in their stable phases and contain holes (empty spaces).*

It is helpful to compare premelting with its more familiar counterpart, the bulk liquid-solid transition. In the latter scenario, the supercooling/superheating phases turn into stable phases by a nucleation process: nuclei (i.e., small regions) of the stable phase are randomly generated by thermal fluctuations, and nuclei large enough can grow continuously into bulk [80]. In the premelting transition, the liquid regions in the 2nd bilayer (Fig. 3(b)) may be similarly considered as nuclei. The difference, however, is manifested by considering two bilayers together: nuclei are much more likely to form near or under holes in the 1st bilayer, instead of being uniformly distributed.

## Transition dynamics: freezing

A typical trajectory of the freezing direction, where QLL turns into the solid surface, is illustrated in Fig. 5. Similarly, holes are found in the 1st bilayer of QLL (Fig. 5(a), circled in a1-a2; also see Fig. 4(b)). In the beginning, the 2nd bilayer is mainly composed of liquid, with solid islands of varied sizes straggling in it (b1). After some time, large solid islands appear in the 2nd bilayer where holes in the 1st bilayer locate (b2, a2, circled). At first glance, this seems to be an analog to the premelting direction discussed earlier (Fig. 3): holes in the 1st bilayer help the formation of solid nuclei (i.e., the large solid islands) and promote the phase transition. However, the large nuclei under holes do not grow further in this case. Indeed, it is another nucleus away from holes that finally grows larger (a3, b3, circled), and spreads into the whole surface (a4, b4). And it is not a coincidence either: though large nuclei can form under holes, they are unlikely to induce freezing (Supplemental Material Sec. 5 [60]). This is different from its bulk counterpart, where the nuclei size is the key factor in triggering phase transitions. Indeed, nuclei under holes only induce freezing in one of the five independent runs, and it takes a rather long time (Fig. S7 [60]).

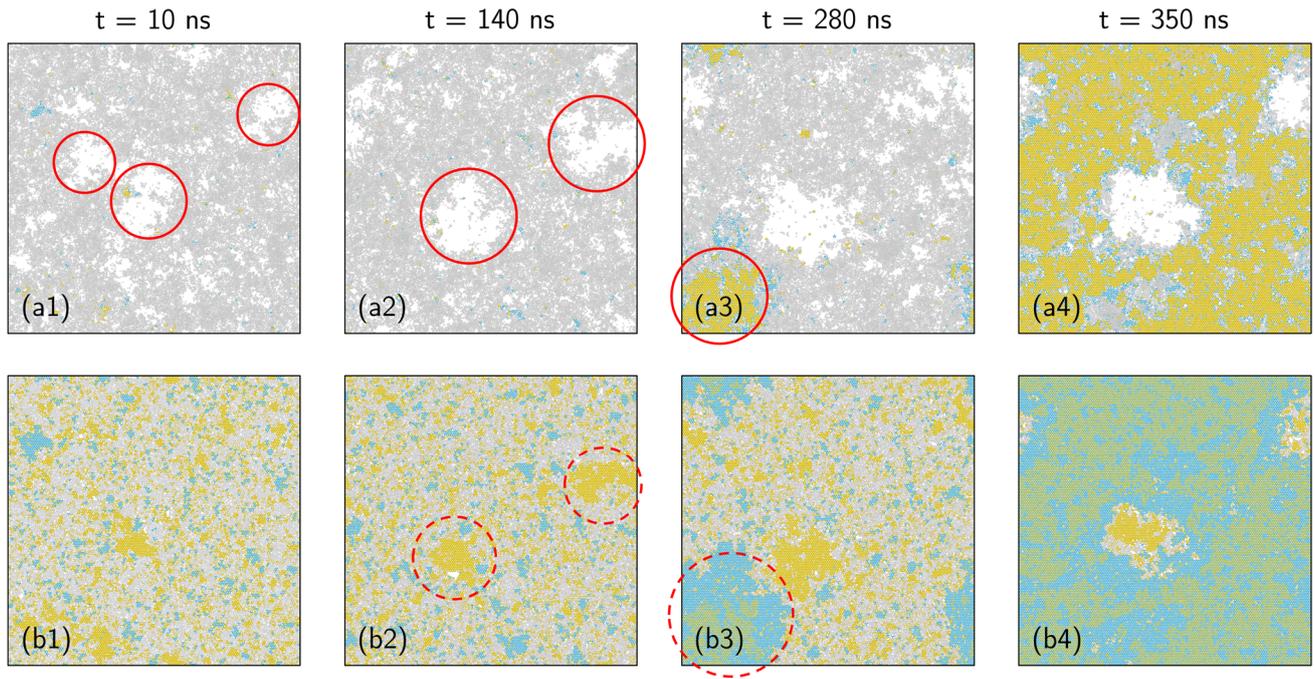

*Figure 5. The structure of ice surface during a premelted -> solid transition. Colors represent different structural types (see legends in Fig. 1). The system size is ~70*70 nm. The simulation is performed at 293.3 K, starting from a premelted surface. (a1~a4) the 1st bilayer, at 10/140/280/350 ns of the simulation. (b1~b4) the 2nd bilayer. Some holes and solid nuclei are circled (see main text).*

Such observation reveals the different roles of holes during surface melting and freezing: though holes exist in both scenarios, they seem to only promote the transition to premelted phases, but not reversely. From another perspective, though large nuclei under holes exist in both scenarios, they only significantly contribute to premelting but not freezing. This contrast probably traces back to the origin of holes on ice surfaces: the molecule redistribution between bilayers. Namely, molecules in the 1st bilayer move into lower and upper layers during premelting (Supplemental Material Sec. 6 [60]), resulting in lower molecule density thus growing holes (Supplemental Material Sec. 7 [60]). Growing holes allow the nuclei under holes to grow, which may further trigger the phase transition. In the freezing process, however, holes tend to shrink instead (Supplemental Material Sec. 7 [60]), confining the nuclei under them and preventing them from spreading into the whole surface. Therefore, holes do not help the transition to the freezing phase.

Another notable thing is that the solid grown from the supercooled QLL is partially stacking disordered ice (b3, b4, cubic molecules shown as blue). This is observed in 2 out of 5 runs, indicating that such stacking disorders might also exist in real ice surfaces frozen from QLL. This hypothesis is further supported by the belief that stacking disordered ice is kinetically favored during crystallization [81]. Furthermore, molecules from different nuclei (e.g. under holes) may have different stacking orders within the 2nd bilayer (b3), because of their different origin. As a result, the boundary region between stacking orders remains liquid to the end of simulation (b4) due to lattice mismatch. The lifetime of such boundaries is unknown yet, but they might survive longer if the holes in the 1st bilayer are large, which could result from holes merging to reduce their "edge" energy in a longer timescale.

## Discussion

So far, we have shown the surface inhomogeneity of QLL and its implication for the transition dynamics. In real scenarios, the ice surface may undergo growth or evaporation. Such processes can lead to an ice surface not being terminated with one full bilayer, altering the surface structure including the inhomogeneity. Specifically, excess molecules beyond one bilayer can "fill" the holes and ultimately lead to raised clusters on the surface (Supplemental Material Sec. 8 [60]). On a sufficiently large spatial and temporal scale, these holes or clusters may grow larger and go beyond a single bilayer, resulting in macroscopic inhomogeneity on ice surfaces. Similar inhomogeneity is already observed in experiments with optical microscopy [30], and is suggested in a continuous model [28]. It is desirable that such macroscopic feature (usually at ~10 μm or larger) can be directly generated from the molecular level.

In this study we used the ML-mW water model, which shows a $1^{st}$-order premelting transition at several Kelvins below the melting point (Fig. 1). Though this model can describe many properties of water with good accuracy [56], it cannot be perfect due to the empirical and coarse-grained nature. For example, the diffusion coefficient of water reported by the ML-mW model is ~2× the experimental value [56]. Therefore, the conclusion of this work should not be considered as definitive. Indeed, models like the original mW show a completely continuous premelting transition [21, 45], and the premelting onset reported by different simulations can differ by dozens of Kelvins [11, 15, 21, 25-26, 45]. Therefore it is not surprising if other water models have different premelting dynamics, not to mention experiments with highly variable conditions. However, currently it is difficult to remove this ambiguity, as we have to rely on empirical classical force fields for computational efficiency. These force fields are usually built with the main focus on bulk properties [56-57, 64], making them less reliable for premelting studies. To move towards a more definitive conclusion, an efficient force field of water that can accurately represent the ice surface is desired. Nevertheless, we still feel it helpful to provide a potential picture of the transition dynamics here, indicating the importance of metastable phases, and showing how QLL may behave under a scaling not frequently reached.

## Conclusion

In this work, we have provided a molecular-level picture of the transition dynamics between the premelted and solid ice surfaces. Supercooling and superheating phases exist in the premelting transition, and generally the supercooled phase has smaller fluctuation. The transition between phases roughly follows the nucleation picture, and holes in ice surfaces promote nuclei generation. Nuclei induced by holes can grow and lead to a full transition to the premelting phase, but not to the inverse direction. Sometimes the freezing process leads to the coexistence of different stacking orders in the same bilayer, due to having multiple nuclei concurrently. Further studies may focus on developing water models specifically target ice surfaces, and re-evaluating the premelting behavior of ice including transition dynamics.

## Acknowledgments

This work was supported by the National Natural Science Foundation of China under Grant No. 11974024; we thank Prof. Ji Chen for providing computational resources.

## Appendix

### A. Existence of complete premelting in several water models

In this section, we will compare the stable surface structures of three water models (mW, ML-mW, TIP4P/Ice) near the melting point. To this end, a 10*10 nm perfect ice surface is built for each of these three models, and they are simulated near the melting point of each model for 10 ns. The results are shown in Fig. 6.

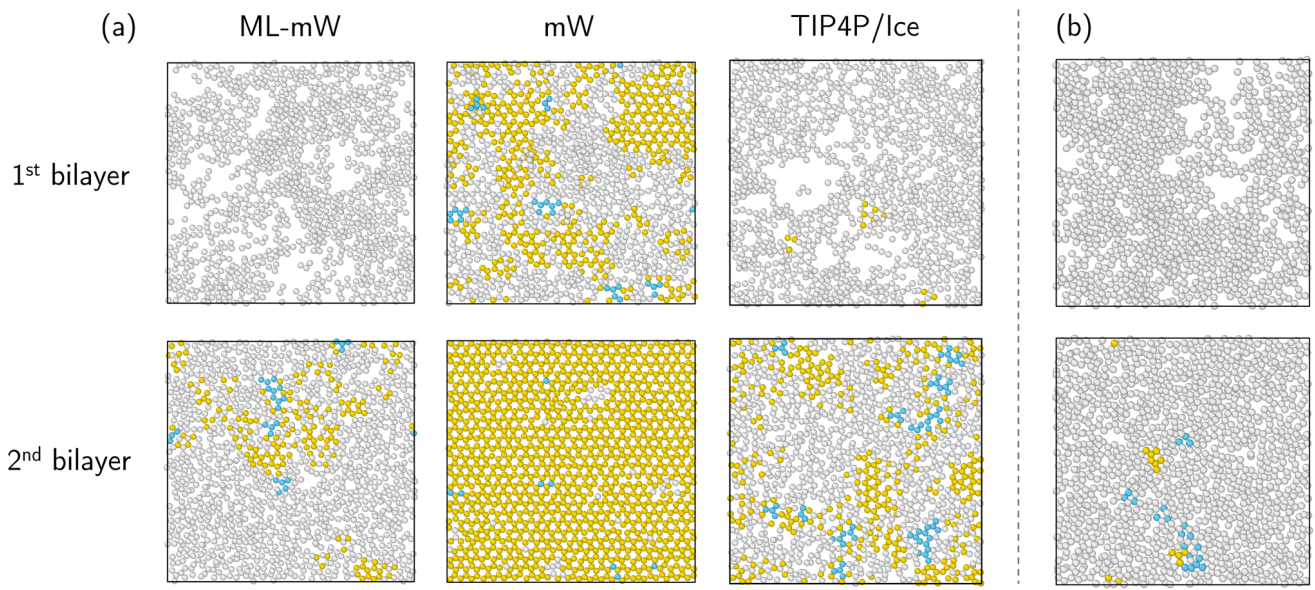

*Figure 6. Ice surface near melting point in three water models, sliced for the 1st and 2nd bilayer. Colors represent structural types (see Fig. 1). Only oxygen atoms are shown for TIP4P/Ice. (a) ice surfaces after 10 ns of simulation each. Temperature: ML-mW -> 295.7K ($T_m$-0.13K), mW -> 275.7K ($T_m$+1K [57]), TIP4P/Ice -> 270K ($T_m$+0.2K [58]). (b) the ice surface of mW model after rapidly throwing molecules and 25 ns of simulation (throw at the rate of 0.2 molecules/fs, with initial velocities of 0.2~0.3 nm/ps towards the surface. 1 bilayer is thrown in total). The ice surfaces keep a similar structure in the 25 ns simulation, i.e., it is at least a metastable phase.*

The ML-mW and TIP4P/ice models share similar features near the melting point: the 1st bilayer is almost completely melted, while the 2nd bilayer is mainly liquid with some solid islands in them (Fig. 6(a)). However, the original mW model is very different: the 1st bilayer is only partially melted at a temperature higher than the melting point, and the 2nd bilayer is still mostly solid (Fig. 6(a)). In other words, the mW ice never completely premelts. To eliminate the effect of potential metastable phases, we tried to artificially make a completely premelted surface for the mW model by rapidly throwing molecules to the surface. It turns out that the mW model could have a completely premelted phase (Fig. 6(b), also see captions), but coexistence tests reveal that it is only metastable even above the melting point (not shown here). Therefore, the mW model does not seem to have a thermodynamically stable complete premelting, while two other models have.

Such difference points out the importance of force field choice in premelting studies. However, note that it does not mean ML-mW and TIP4P/Ice are completely the same in premelting behavior. Indeed, the TIP4P/Ice model is reported to show premelting down to 200K [10] (also see Supplemental Material Sec. 3 [60]), while the ML-mW shows that only at several Kelvins below the melting point. Nevertheless, the comparison indicates that the ML-mW model is a reasonable choice for premelting studies, especially when the large scaling of simulations prohibits using full-atomic models.

## B. Order parameter

The molecules in ice surfaces are categorized by their local structures through this work, by the method described in Appendix A of ref. [72]. This method is originally purposed for the Ge/Si system but is also applicable for systems with similar lattices, such as ice [35]. It exploits the fact that the $I_h$ ice consists of two interleaved HCP lattices, and the $I_c$ ice consists of two interleaved FCC lattices; therefore, we can determine the structure type of a molecule by checking out if that molecule and its *second* neighbors form an FCC/HCP lattice. The latter task can be done by the widely-used common neighbor analysis (CNA) [82], which works by classifying neighbor molecule pairs by their local environments; readers are redirected to ref. [82] for details of CNA. The procedure of the whole analysis is summarized below:

(1) Select a molecule ("central molecule") to be identified;
(2) Find 4 nearest neighbors of the central molecule ("1$^{st}$ neighbors");
(3) Find 4 nearest neighbors for each of the 1$^{st}$ neighbors ("2$^{nd}$ neighbors"). This should give 12 molecules in total, excluding the central molecule itself (which is counted 4 times);
(4) Perform CNA on the central molecule and 2$^{nd}$ neighbors. The cutoff radius of CNA is $r_{CNA} = r_0(1 + \sqrt{2})/2$, where $r_0$ is the average distance of twelve 2$^{nd}$ neighbors from the central molecule. For the central molecule of $I_c$ ice, all center-neighbor molecule pairs are of the 421 type; while for $I_h$ ice, half of them are 421 and half are 422 [82].

By repeatedly performing the procedure above, all water molecules can be categorized into the following groups:

(a) A central molecule of $I_h$ ice;
(b) A central molecule of $I_c$ ice;
(c) Does not belong to (a) or (b), but is a 1$^{st}$ neighbor of a group (a) molecule;
(d) Does not belong to (a) or (b), but is a 1$^{st}$ neighbor of a group (b) molecule;
(e) Does not belong to (a)~(d), but is a 2$^{nd}$ neighbor of a group (a) molecule;
(f) Does not belong to (a)~(d), but is a 2$^{nd}$ neighbor of a group (b) molecule;
(g) None of the above.

For our purpose, the structure type "hexagonal ($I_h$)" corresponds to group (a)(c)(e), the "cubic ($I_c$)" corresponds to (b)(d)(f), and the "liquid" corresponds to (g). Note that it is correct to consider group (c)~(f) as solid molecules: these molecules themselves are "on-site" though some of their neighbors are not. This ensures the surface molecules are correctly categorized (i.e. a perfect ice surface has an order parameter of 1). For the three-atom TIP4P/Ice model, only oxygen atoms are considered during the identification.

It is possible for a molecule to be eligible for both (c) and (d), or both (e) and (f). Usually this occurs at the boundary of $I_h$ and $I_c$ ice, where the local structure may be equivalently interpreted as either. Anyway, the order parameter itself is always well-defined.

Note that this approach is mostly based on geometrical arguments, so there is no need for an arbitrary number to separate two phases. In addition, it is also more robust at solid-liquid interfaces compared to some other choices. Such scenarios occur in the bilayer right under QLL, e. g. the 3$^{rd}$ bilayer in the premelted surface. Fig. 7 shows a typical structure of the 3$^{rd}$ bilayer at 295.2K, with the first 2 bilayers premelted. The bilayer largely keeps the hexagonal structure with only occasional defects, indicating that the order parameter should be close to its maximum. Tabl. 1 shows the order parameter reported by three different definitions. The one used in this work reports ~2.1% of melting, which is reasonable considering the existence of defects. However, the two other choices report much higher percentages.

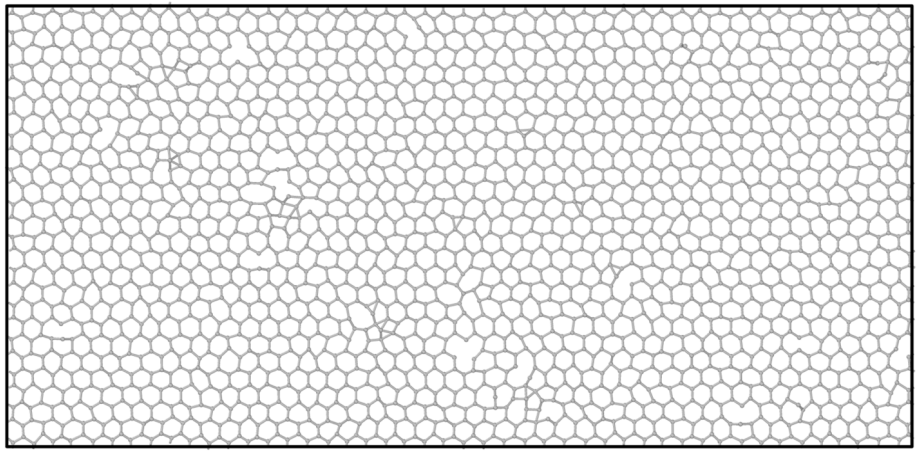

*Figure 7. A typical structure of the 3$^{rd}$ bilayer at 295.2K, with the first 2 bilayers premelted. The system size is ~20 x 10 nm. Bonds are drawn for visualization only; they are not used in analysis.*

|  | This work [72] | Local tetrahedrality [83] | Local Steinhardt $q_6$ [84] |
| --- | --- | --- | --- |
| Value | 0.9787 | 0.8872 | 0.5390 |
| Possible range | [~0.005, 1] | [~0.65, ~0.97] | [~0.25, ~0.74] |
| Percentile (% melted) | ~2.1 | ~26 | ~41 |

*Table 1. The order parameter under three definitions, and the corresponding percentile in their possible ranges. The lower limit is estimated by the 1$^{st}$ bilayer of the premelted surface, and the upper limit is estimated by the innermost movable (8$^{th}$) bilayer of the system. The percentile may be interpreted as how much of the bilayer has melted (0%=as melted as the 1$^{st}$ bilayer, 100%=as solid as the innermost bilayer). See captions of Fig. 8 for definitions of the last two order parameters.*

Fig. 8 recalculates the static premelting profile (Fig. 1(a)) under several order parameters. The exact behavior differs, but all of them suggest a 1$^{st}$-order type transition at the same temperature and a QLL of ~2 bilayers, agreeing with Fig. 1(a).

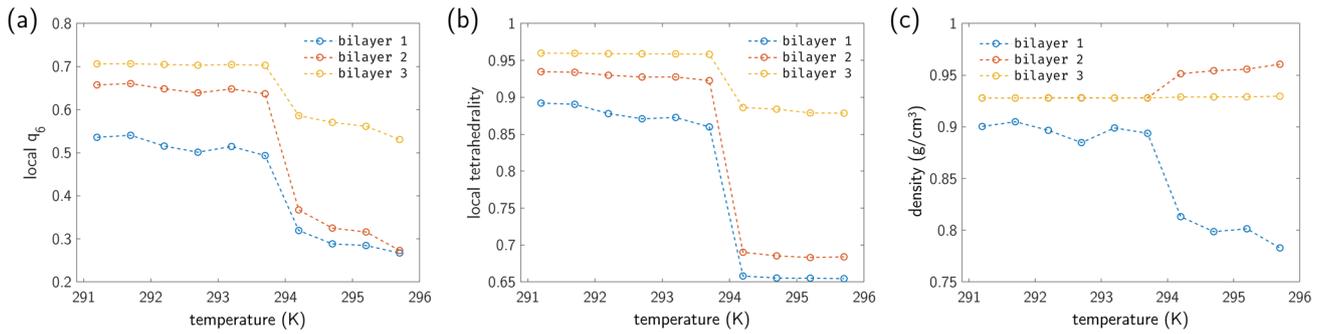

*Figure 8. The static premelting profile in Fig. 1(a) under several order parameters.*

(a) the local Steinhardt parameter $q_6$, defined for each molecule $i$ as $\frac{1}{n}\sum_j [\boldsymbol{q}_6(i) \cdot \boldsymbol{q}_6(j)]$, where $\boldsymbol{q}_6$ is a 13-vector $(q_{6(-6)}, q_{6(-5)}, \ldots, q_{6(6)})$ defined as $q_{6m}(i) = \frac{1}{n}\sum_j Y_{6m}(\boldsymbol{r}_{ij})$. In these expressions, $n$ is the number of molecules in the first coordination sphere (<0.34 nm) of $i$, the summation over $j$ includes all molecules in that sphere, $\boldsymbol{r}_{ij}$ is the distance vector between $i$ and $j$, $Y_{6m}$ are sixth order spherical harmonics. This parameter is calculated using PLUMED [84].

(b) the local tetrahedrality [83], defined for each molecule $i$ as $\left(1 - \frac{9}{2n(n-1)}\sum_{\langle j,k \rangle}\left(\cos\theta_{jik} + \frac{1}{3}\right)^2\right)$, where $\theta_{jik}$ is the angle formed by molecules $j/i/k$, and the summation over $j,k$ includes all pairs of molecules in the first coordination sphere of $i$. Note that the "unnormalized" tetrahedrality [85], where the coordination number $n$ is always 4, is not a good choice for premelting studies since molecules are under-coordinated on surfaces.

(c) the density, defined for each bilayer as $\frac{NM}{Ah}$, where $N$ is the number of molecules in the bilayer, $M$ is the mass of a water molecule, $A$ is the surface area of the system, and $h$ is the height of a single bilayer (0.73487/2 nm). The opposite trend of density change in the first 2 bilayers reflects the molecule redistribution during premelting (Supplemental Material Sec. 6 [60]).

# Supplemental Material

For the article: *Transition dynamics and metastable states during premelting and freezing of ice surfaces*

Shifan Cui*[a], Haoxiang Chen[b]

[a]International Center for Quantum Materials, School of Physics, Peking University, 209 Chengfu Road, Haidian District, Beijing 100871, China.
[b]School of Physics, Peking University, 209 Chengfu Road, Haidian District, Beijing 100871, China.
*academic_sfc@outlook.com

# Methods

Some of the simulation setup has been mentioned in the main text; this section will provide additional details.

## ML-mW

The ML-mW model is used in most simulations in this work. The lattice parameter is determined by an NPT simulation at 294 K, resulting in a=b=0.45015 nm, c=0.73487 nm. The variation of lattice parameter due to temperature change is < 0.0001 nm in the temperature range studied in this paper and is neglected. All simulations are performed in the NVT ensemble unless otherwise noted, using the Nose-Hoover thermostat[1-2] with a relaxation time of 1.25 ps. The timestep for MD is 5 fs. A slab with 10 bilayers is used for simulating the ice surface. The bottom 2 bilayers are fixed during the simulation (by excluding them from time integration and thermostating), and the top 8 bilayers are movable. A reflective layer is placed above the ice surface to avoid losing molecules. A vacuum layer exists between the ice surface and reflective layer, with thickness >> 0.4 nm ( ≈ the cutoff of ML-mW model). Periodic boundary conditions are imposed in two other directions.

## The original mW

The original mW model is used in Appendix A. Most of the simulation setup is same as ML-mW except for the force field parameters. The lattice parameter used for mW model is a=b=0.4427 nm, c=0.72028 nm[3].

## TIP4P/Ice

The TIP4P/Ice model[4] is used in Appendix A and Supplemental Sec. 3. The ice structures are generated by the GenIce2 software[5-6] with a density of 0.90634 g/cm$^3$. GenIce2 generates ice structures with stochastic arrangements of hydrogen atoms and zero net dipole moment. A slab with 10 bilayers is used for simulating the ice surface. No molecules are fixed in the simulation (i.e., it is a double-sided slab). All reported statistics about the ice surface are averaged for both sides. Periodic boundary conditions are imposed in all directions. A large vacuum layer (~3x thickness of the slab) is placed "between" two sides to reduce interactions between periodic images. Note that whole molecules should not be split apart when adding the vacuum layer. The velocity-Verlet integrator and Nose-Hoover thermostat are used for being consistent with LAMMPS. All simulations are performed in the NVT ensemble with a thermostat relaxation time of 1.25 ps. The timestep for MD is 2 fs. The coulomb part of interaction is calculated by the fast smooth particle-mesh Ewald method, with the short-range part cut at 0.9 nm. The van der Waals part is calculated with a plain cut-off of 0.9 nm, with long-range corrections for energy and pressure. The linear center-of-mass velocity is removed every 100 steps to mitigate the drifting of slab.

## Getting the stable phase

In the main text, we showed the existence of supercooling/superheating in premelting. They introduce some difficulty in finding the (thermodynamic) stable phase, since these metastable phases may persist in a finite-time simulation. To find the stable phase in such scenarios, we used a scheme similar to the direct-coexistence technique for finding bulk melting points. First, the premelted and solid ice are prepared independently in advance, by NVT simulations in corresponding temperature ranges (e.g., 295 and 293 K).

After that, they are concatenated together (Fig. S1):

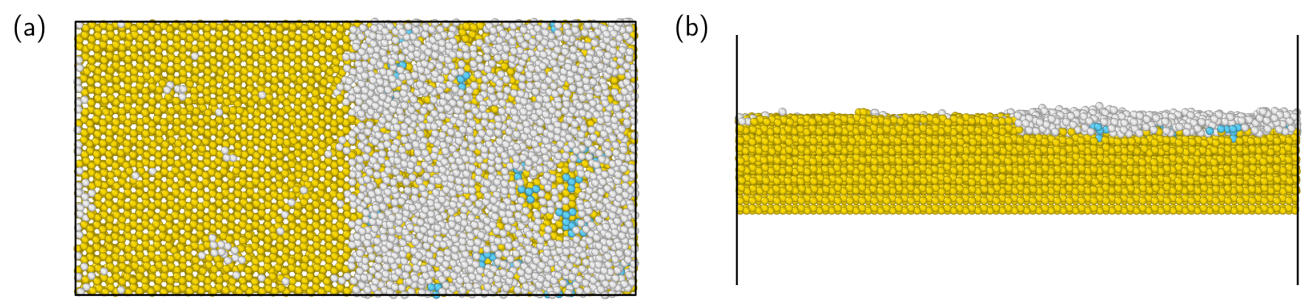

*Figure S1. The concatenation of premelted (right) and solid (left) ice, after relaxation. (a) top view, (b) side view. Colors represent structural types (see Fig. 1, main text). The total system size in this illustration is 20*10 nm, and each section spreads for 10*10 nm. Other sizes may be used in this study too (see text). The relaxation has an imperceptible effect on appearance (it is only meant to separate overlapping molecules).*

The obtained structure is slightly relaxed to avoid molecule overlap, then the regular NVT simulation is performed. Note that the simulations are performed at constant volume (instead of constant pressure in its bulk version), since the surface does not affect lattice parameters of bulk ice (because it only takes up an infinitesimal portion of molecules in the real world, where the "slab" contains many bilayers). At the thermodynamic limit, a certain phase will dominate the surface after sufficient time, telling us the stable phase at this temperature. In finite-size simulations, the dominating phase may vary in a narrow temperature range and an estimate of the premelting onset is derived. This scheme is used to get results in Fig. 1(a) (with a total lattice size of 20*10 nm) and Supplemental Sec. 2 (total lattice size of 40*20 nm).

### Splitting bilayers

In this work, we split the ice surface into multiple bilayers to analyze its order and structure. For the mW and ML-mW models, this can be done by counting from the fixed layers at bottom. Namely, the bisector of two bottom layers is found first, and other bisectors are derived by successively adding the height of one bilayer. For the TIP4P/Ice model, the bisector of two middle layers is found first instead, by averaging the z coordinate of all oxygen atoms. In this approach, all bilayers except the top two have almost same number of molecules, and the bisector of top two bilayers is very close to the local minimum of molecule distribution histogram in z direction. The change of molecule numbers in top two layers is induced by premelting (Supplemental Sec. 6).

# 1 The melting point of ML-mW model

In the original paper of ML-mW model, the melting point is reported as 289K. To study its premelting behavior, we re-determined the melting point with higher accuracy using the direct-coexistence technique. The model used in direct-coexistence simulations is ~7.2*5.5*13.2 nm in size and has ~16000 molecules in total, half filled by water and half filled by ice. A large model reduces stochastic error, and also reduces systematic error caused by the difference between NPT and the exact $NP_zT$ ensemble[7-8]. 5 independent runs (with different random seeds for initial velocities) are performed for each temperature under the NPT ensemble. Each run lasts for 200 ns, or until a certain phase dominates the system. The results are summarized in Tabl. S1. The estimated melting point is 295.83±0.5K.

| Temperature/K | 295 | 295.33 | 295.67 | 296 | 296.33 |
|---|---|---|---|---|---|
| **Solid** | 5 | 5 | 1 | 0 | 0 |
| **Liquid** | 0 | 0 | 0 | 0 | 5 |
| **Undetermined** | 0 | 0 | 4 | 5 | 0 |

*Table S1. The result of direct-coexistence tests for the ML-mW model. "Undetermined" means neither phase dominated the system after 200 ns of simulation.*

## 2 The premelting onset of ML-mW model

The premelting onset of ML-mW model is determined by the coexistence technique demonstrated in the Methods section (Fig. S1), with total system sizes of 40*20 nm. 4 independent runs with different random seeds for initial velocities are performed for each temperature. Each run lasts for 75~235 ns. The results are summarized in Tabl. S2; the estimated transition temperature is 293.85±0.15K. This is shown as the green shadow in Fig. 1(a) and 2(a) of main text.

| Temperature / K | 293.7 | 293.8 | 293.9 | 294.0 |
|---|---|---|---|---|
| **Solid** | 4 | 1 | 1 | 0 |
| **Premelted** | 0 | 1 | 2 | 4 |
| **Undetermined** | 0 | 2 | 1 | 0 |

*Table S2. The results of premelting coexistence tests. "Undetermined" means neither phase dominates the system at the end of simulations.*

## 3 Slow transition dynamics of ice surfaces under low temperatures

One uncertainty of determining premelting onset comes from the long-surviving non-equilibrium states. In the ML-mW model, it mostly refers to metastable phases: both supercooling and superheating phases can be observed in a small temperature range (293.3~294.3 K). If the temperature is lower, larger uncertainty may arise from the slow transition dynamics, since the molecular diffusion is suppressed. To demonstrate this, here we consider the TIP4P/Ice model with a much lower premelting onset. Simulations are performed starting from either perfect or premelted surface at three temperatures: 180, 200, and 220 K. Each simulation is performed in a 10*10 nm simulation cell and lasts for 100 ns. Using a relatively large cell of 10*10 nm (compared to ~3*3 nm in a previous work[9]) reduces finite-size effects and fluctuations, and is necessary to reproduce the slow surface dynamics.

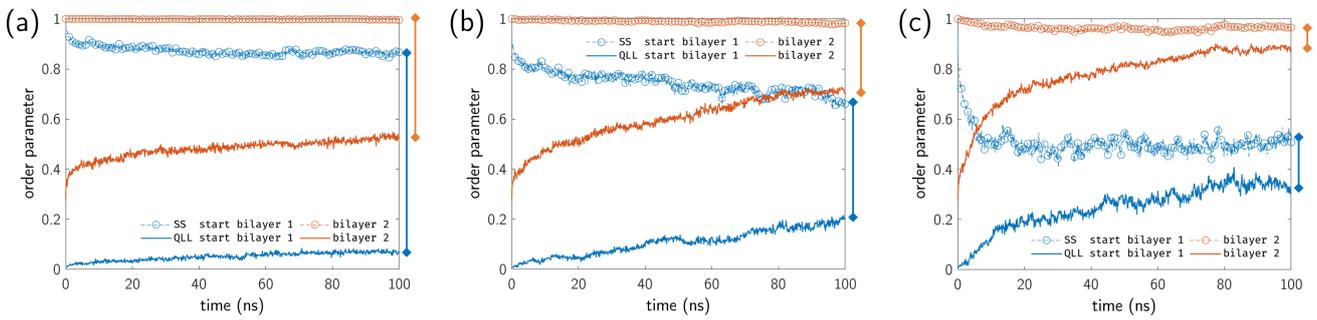

*Figure S2. The order parameters of first 2 bilayers, during 100 ns simulations at (a) 180 K (b) 200 K (c) 220 K. SS=solid surface. Vertical lines on the right indicate the difference in order parameters between supercooled and superheated phases at the end of simulations.*

Fig. S2 shows the evolution of surface order during the simulations. Notably, the liquid and solid phases do not reach the same state during the simulation: there is a non-zero difference between order parameters of the same bilayer at the end (vertical lines on the right). Since there is only one global equilibrium state, at least one of them (more likely both) has not reached equilibrium after 100 ns. In other words, non-equilibrium states can survive in a wide range of temperatures for a considerable time. The difference is larger when the temperature is lower, indicating that the transition is kinetically limited. Indeed, we found that the coexistence technique for premelting (see Methods section) does not work well at such temperatures because the kinetics are so forbidding that a boundary does not help much.

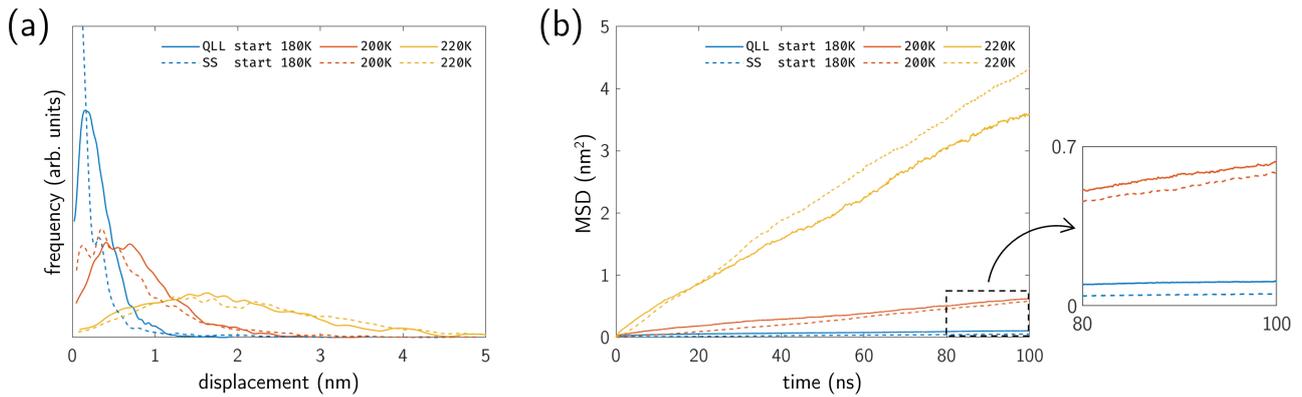

*Figure S3. Surface diffusion in the 1st bilayer. Only horizontal displacement is considered, and the overall drift of bulk is removed beforehand. (a) the distribution of molecule displacement during the 100 ns simulations. The frequency is normalized so that the total area under each curve is same. Curves are slightly smoothed for better appearance. (b) the mean squared displacement (MSD) as function of time. The framed region at the bottom-right corner is magnified on the right. The diffusion coefficients are proportional to the slopes at equilibrium (not really achievable here, but can be approximated by the ending part of simulations).*

The effect of low temperatures can be better understood by checking the surface diffusion. Fig. S3(a) shows the total horizontal displacement of molecules in the 1st bilayer in simulations. The QLL and solid phases have similar profiles for molecule displacements, and the QLL phase diffuses only slightly faster. Indeed, the diffusion of both phases is quite slow, especially at low temperatures (180 and 200 K) where the molecule displacement is comparable to the lattice parameters. In other words, the "QLL" behaves like a solid glass for diffusion at these temperatures (where two phases do not merge yet). The mean square displacement (MSD) of molecules demonstrates this further (Fig. S3(b)): the difference in diffusion coefficient between two phases is minor and much less than that caused by temperature change. The slow diffusion means neither of the two phases is kinetically eager to convert into the other, so either of them may exist under these temperatures. It also implies that the surface structure is not everything for premelting: certain physical properties, like the diffusion coefficient, may have no direct relation to the structure. In other words, multiple premelting "onset" may exist even for the same force field, depending on the physical properties of interest.

# 4 Effect of holes on the distribution of liquid molecules

In the main text, we pointed out that the liquid molecules in the 2nd bilayer are more likely to appear near or under holes (Fig. 3, main text). To demonstrate this more generally, we checked the distribution of horizontal distance between liquid molecules in the 2nd bilayer and nearest holes in the 1st bilayer. This is shown in Fig. S4. Compared to a random position (orange bars), liquid molecules in the 2nd bilayer are much more likely to appear within or near a hole (blue bars).

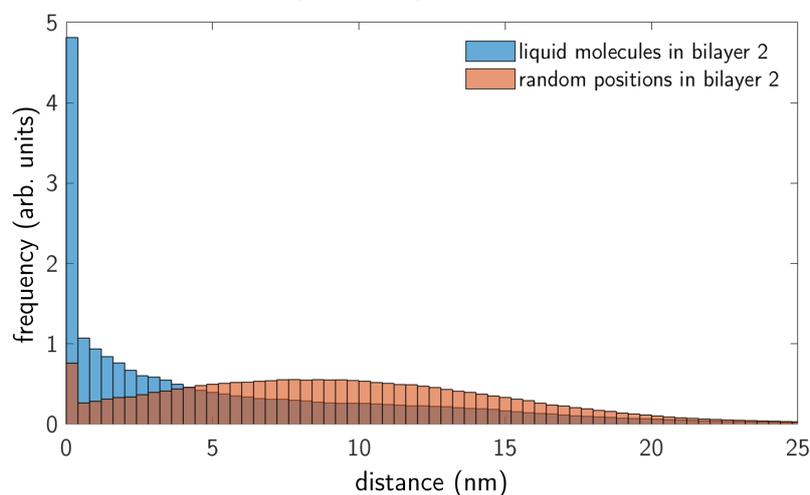

*Figure S4. Histogram of horizontal distance between liquid molecules and nearest holes. The statistics are averaged over five trajectories, from the beginning of simulations to the point before phase transitions actually start. For comparison, the distance distribution between a random position and its nearest hole is also shown.*

*To identify holes in bilayer 1, the molecules are first projected into the XY plane, then an alpha shape[10] is constructed from them with α=0.35 nm. Holes with area < 4 nm² are filled, and the boundary of the alpha shape gives the boundary of remaining holes. Only molecules belonging to the largest cluster in bilayer 1 are considered in the whole process, where two molecules belong to the same cluster if their distance is smaller than 0.35 nm. This is intended to avoid interference from scattered molecules in holes. The reported distance is the minimum distance between a molecule in bilayer 2 and a boundary molecule of the alpha shape (considering periodic boundary conditions). Molecules located in a hole (after being projected to the XY plane) are reported as distance=0 and are counted into the first bar from the left.*

The increasing appearance of liquid molecules supports the formation of nuclei near holes. It may be helpful to look at a single trajectory as an example. Fig. S5(a) shows the frequency of liquid molecules appearing in the 2nd bilayer after holes have formed and before the phase transition actually starts, in the trajectory corresponding to Fig. 3 of main text. Except for the nuclei finally triggering transition (the solid white circle), several other nuclei are also found (dashed white circles). Fig. S5(b)~(f) show the structure of 1st bilayer at several typical time, with significant holes circled out. Notably, most of the holes (solid circles) are found near the location of nuclei. Furthermore, high liquid frequency (in Fig. S5(a)) generally corresponds to large and long-life holes: for example, holes around two regions with highest frequencies in (a) are found in all of the five timepoints. Such correspondence indicates the relation between holes and liquid nuclei during the premelting transition.

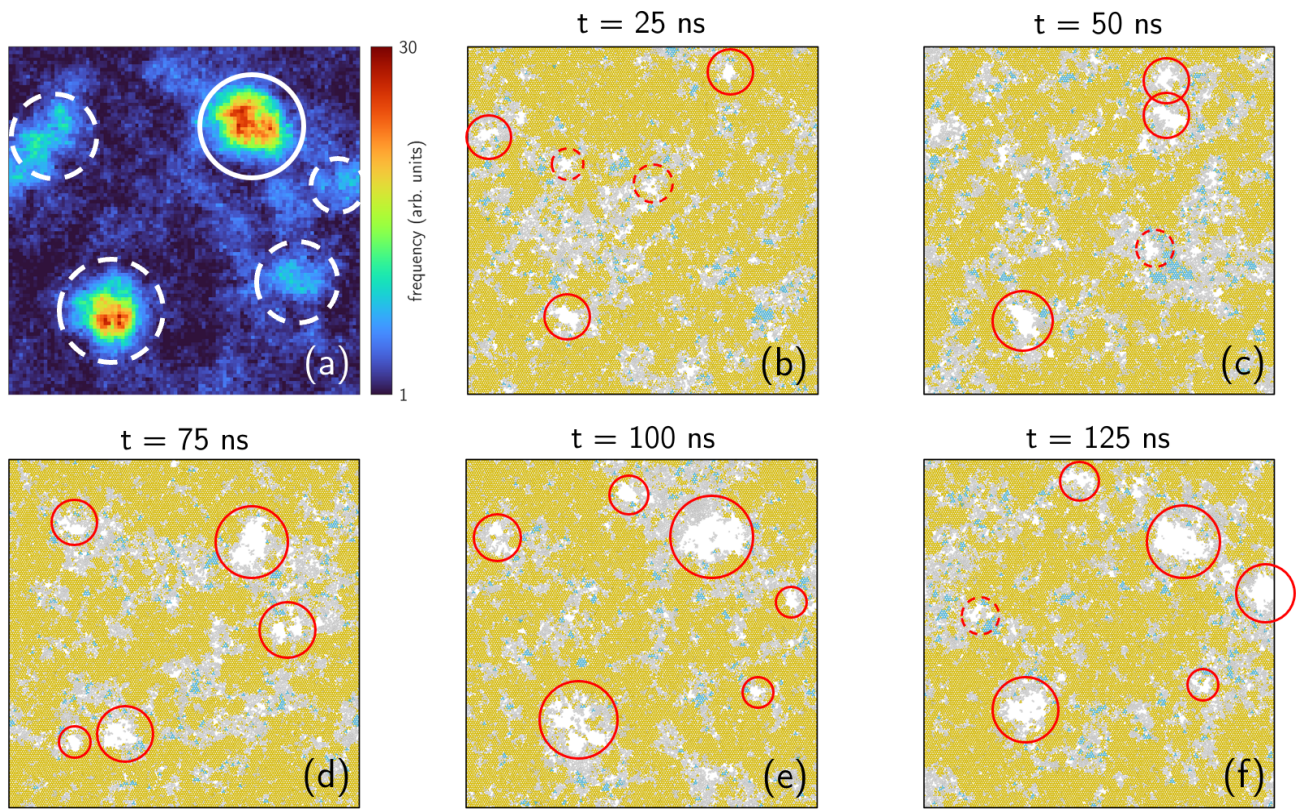

*Figure S5. Relation between holes and liquid nuclei. (a) the frequency distribution of liquid molecules appearing in bilayer 2, during 25~125 ns of the simulation. Some nuclei (regions with high frequencies) are circled out (see text). (b~f) the 1st bilayer at five points during the simulation. Some holes are circled out (see text). Colors represent structural types (see Fig. 1, main text).*

# 5 Solid nuclei under holes are unlikely to induce freezing

In the main text (Fig. 5), we demonstrated an instance where nuclei under holes are large but did not trigger the freeze of QLL. To see if this is a general tendency, we summarized all appearances of nuclei (solid clusters in bilayer 2) during five runs before the phase transition actually starts. Fig. S6(a) shows the histogram of nuclei size distribution, and the proportion of nuclei under holes in each size range. Notably, most of the nuclei are small (~100 molecules or less), and most small nuclei are not located under holes. However, the proportion of nuclei under holes increases significantly with the nuclei size, and reached near 100% for nuclei with >2000 molecules (orange line). In other words, large nuclei are much more likely to appear under holes. Nevertheless, they only induced the transition in 1 out of 5 runs, even though they are much larger. That is to say, nuclei under holes are unlikely to induce freezing.

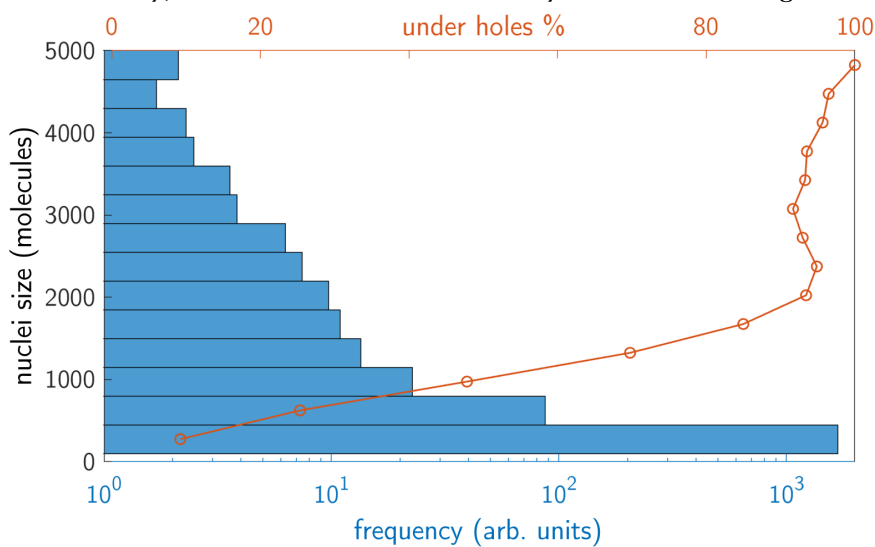

*Figure S6. Statistics of nuclei in supercooled QLL at 293.3 K, averaged over five independent runs. Blue bars: histogram of nuclei sizes; bar lengths are in log scale. The relative frequency (X axis at bottom) roughly represents the nuclei size distribution of the premelted surface during the simulation. Only nuclei with 100~5000 molecules are considered. Orange lines: the percentage of nuclei under holes in each size interval, in linear scale from 0% to 100%.*

*Nuclei are identified by cluster analyses within solid molecules in the 2$^{nd}$ bilayer: solid molecules with distance < 0.35 nm belong to the same nuclei. "If a nucleus is under a hole" is determined by whether molecules in bilayer 1 are found within 1 nm (horizontal distance) around the center of mass of the nuclei. For this purpose, only molecules belonging to the largest cluster in bilayer 1 are considered (see captions of Fig. S4). This crude approach does make some misjudges (e.g. from concave nuclei or occasional small holes on large nuclei) but the accuracy is enough for discussion here.*

It is still possible for a nucleus under a hole to induce the freezing, though it is uncommon. Fig. S7 shows the only instance of such in 5 runs. The transition starts at t≈1300 ns, when a part of the nucleus extends beyond the hole in bilayer 1 (red circles). Note that the elapsed time until observing this transition (1.3μs) is significantly larger than the general MD timescale.

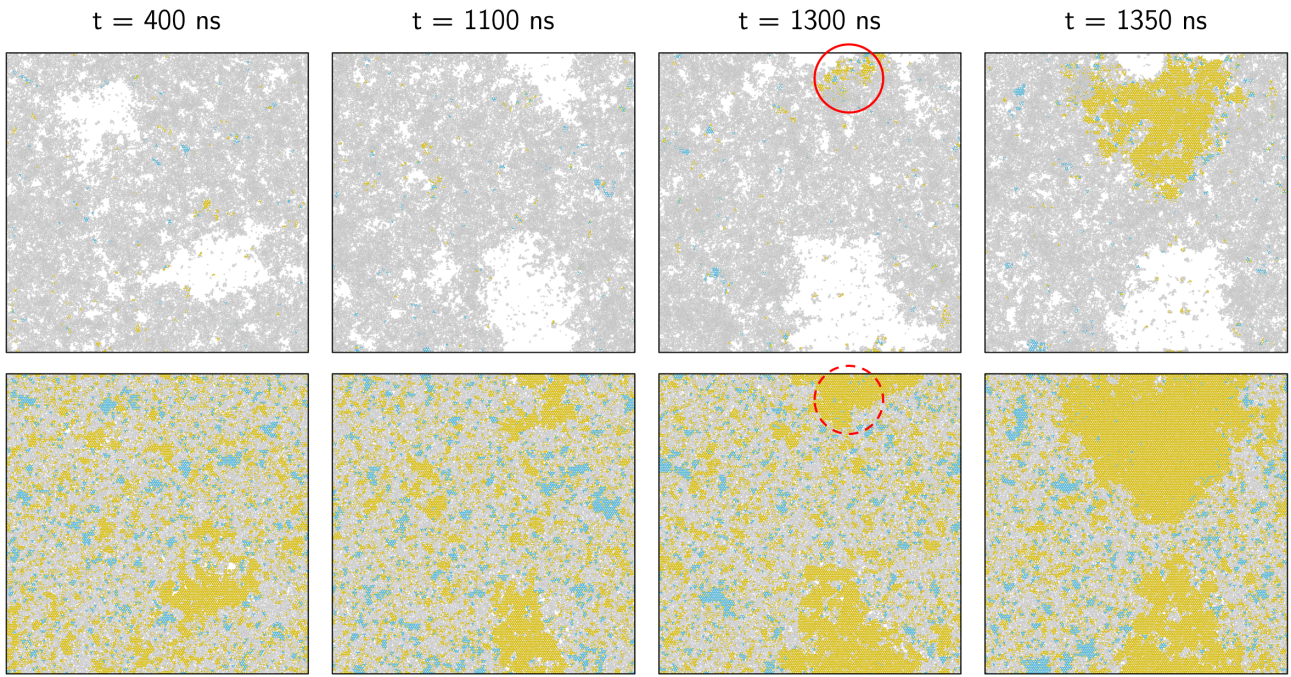

*Figure S7. An instance of surface freezing induced by a nucleus under a hole.*

# 6 Redistribution of molecules between bilayers during premelting

In the main text, we have shown the appearance of holes on the ice surface. Indeed, it can be related to the redistribution of molecules between bilayers on the surface. First we define the "scaled density" of a bilayer, $\rho_s$:

$$\rho_s(X) = \frac{\rho(X)}{\rho_{perfect}}$$

Where $\rho$ is the density of bilayer X, $\rho_{perfect}$ is the density of perfect ice. This scaling makes perfect ice having $\rho_s = 1.0$. An increase of $\rho_s$ for a certain bilayer indicates molecules flowing into that bilayer, and vice versa.

To study the vertical exchange of molecules during premelting, an MD run is performed at four temperatures (250/293.5/295/290 K) successively, with each temperature lasting for 20 ns. The $\rho_s$ of bilayer 0~3 are shown in Fig. S8 ("bilayer 0" means the region above bilayer 1, assuming the bilayer 1 is of one bilayer, i.e., 0.73487 nm thick). At low temperatures (250 K), each bilayer keeps its density (0~20 ns, Fig. S8). The story begins at 293.5 K, where the transition point is not reached yet but the surface already become somewhat disordered (similar to Fig. 1(c)). At this temperature, some molecules start to leave the 1st bilayer (20~40 ns), giving a slight decrease (<5%) of density in bilayer 1. After the premelting transition occurs (45~60 ns), the $\rho_s$ of bilayer 1 drops further (15~20%), and that of bilayer 0 and 2 increase. In other words, the molecules in bilayer 1 move to the upper and lower layers in the premelting transition, and bilayer 1 becomes less dense so holes grow. On the other hand, the molecules in bilayer 0 and 2 go back to bilayer 1 after the surface refreeze (60~80 ns), leading to a denser bilayer 1 and shrinking holes.

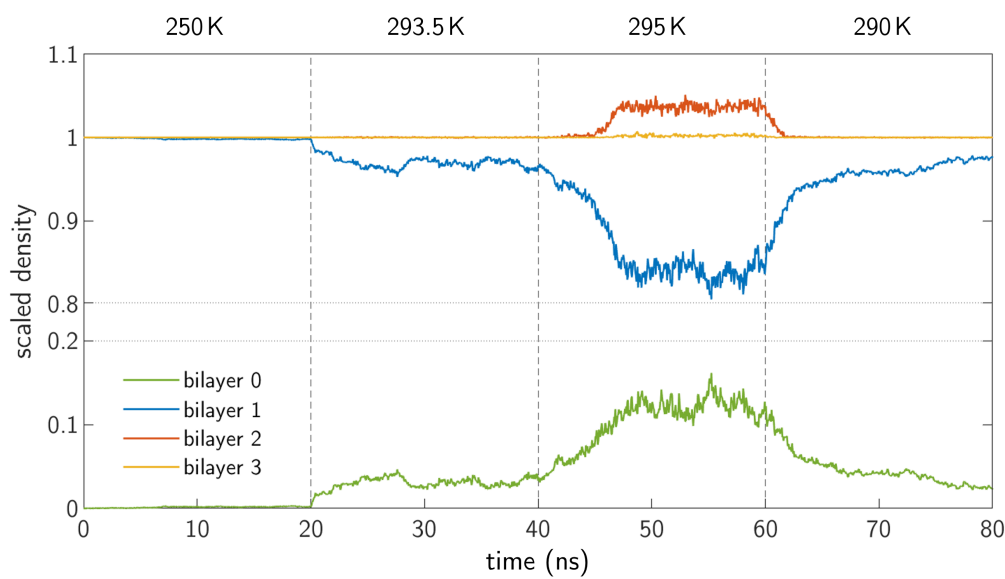

*Figure S8. Scaled density of bilayer 0, 1, 2, and 3 (see text for definition), during an MD simulation at 250, 293.5, 295, and 290 K, successively. The y-axis is broken in the middle. Each temperature lasts for 20 ns (time boundaries are shown as dashed vertical lines). The system size is ~20*20 nm.*

# 7 The trend of hole sizes during premelting and freezing

In Supplemental Sec. 6 we argued that the redistribution of molecules can lead to growing or shrinking holes during the phase transition. This can be made clearer by drawing up the change of hole sizes during the phase transition. Fig. S9(a) shows the trend of the largest hole sizes during the freezing transition in all five trajectories. Notably, the holes tend to shrink after the freezing started for some time (t ≈ 50 ns). Fig. S9(b) shows the trend during the premelting transition, indicating the tendency of holes to grow after the premelting starts.

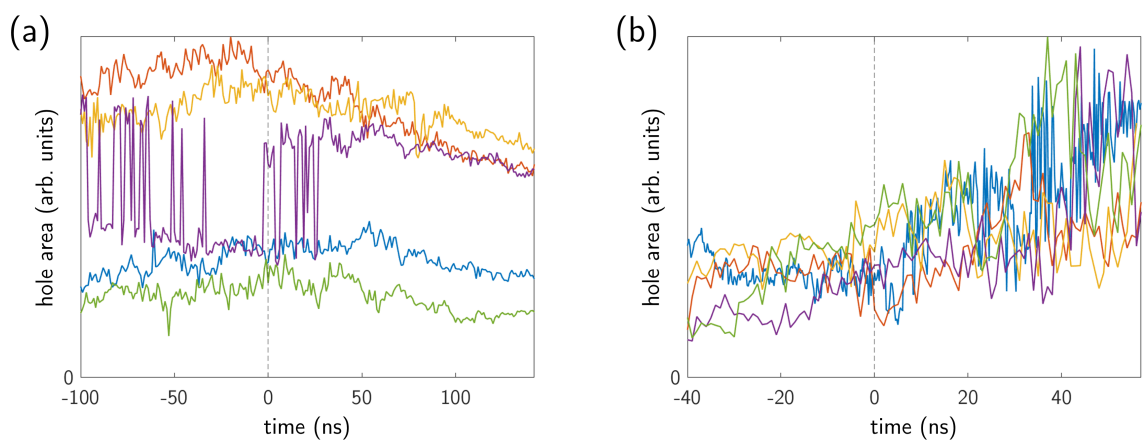

*Figure S9. Sizes of largest holes in the system, during (a) freezing at 293.3 K, and (b) premelting at 294.2 K, in all five simulations. Vertical lines: the starting point of the freezing/premelting transition, estimated by looking at the change of nuclei sizes in trajectories (this introduces some uncertainty but should not affect the discussion here).*

*To measure the size of holes, molecules belonging to the largest cluster in bilayer 1 are extracted (see captions of Fig. S4). Holes are identified as connected regions (considering periodic images) not covered by any molecule in the XY-plane, assuming each molecule has a radius of 0.35 nm. This approach is slightly different from the one in Fig. S4 (due to technical reasons) but should give roughly same results. Large fluctuations in a short time (almost "vertical" in the curves) are irrelevant noises caused by adjacent holes merging and detaching.*

# 8 QLL with excess molecules on the surface

Until now all of our simulations begin from full-bilayer terminated ice surfaces, or systems with same number of molecules as that. In real systems the ice may undergo growth or evaporation, leading to excess molecules beyond one full bilayer on the surface. To study those scenarios, we may introduce excess molecules by throwing molecules toward a full-bilayer terminated surface (see captions of Fig. S10). These excess molecules may fill the holes and lead to clusters on the surface, as shown in Fig. S10 (colored in red). Notably, most of clusters sit on the liquid region of bilayer 1 when the temperature nears the premelting onset (293.5/294 K). This is quite different from clusters observed in the original mW model, where most clusters sat on the solid region[3]. At lower temperatures (293 K), more clusters are found above solid regions. It is yet unclear if such contrast implies a crossover in the relative stability of clusters' locations, or is merely because more solid exists in bilayer 1 at low temperatures.

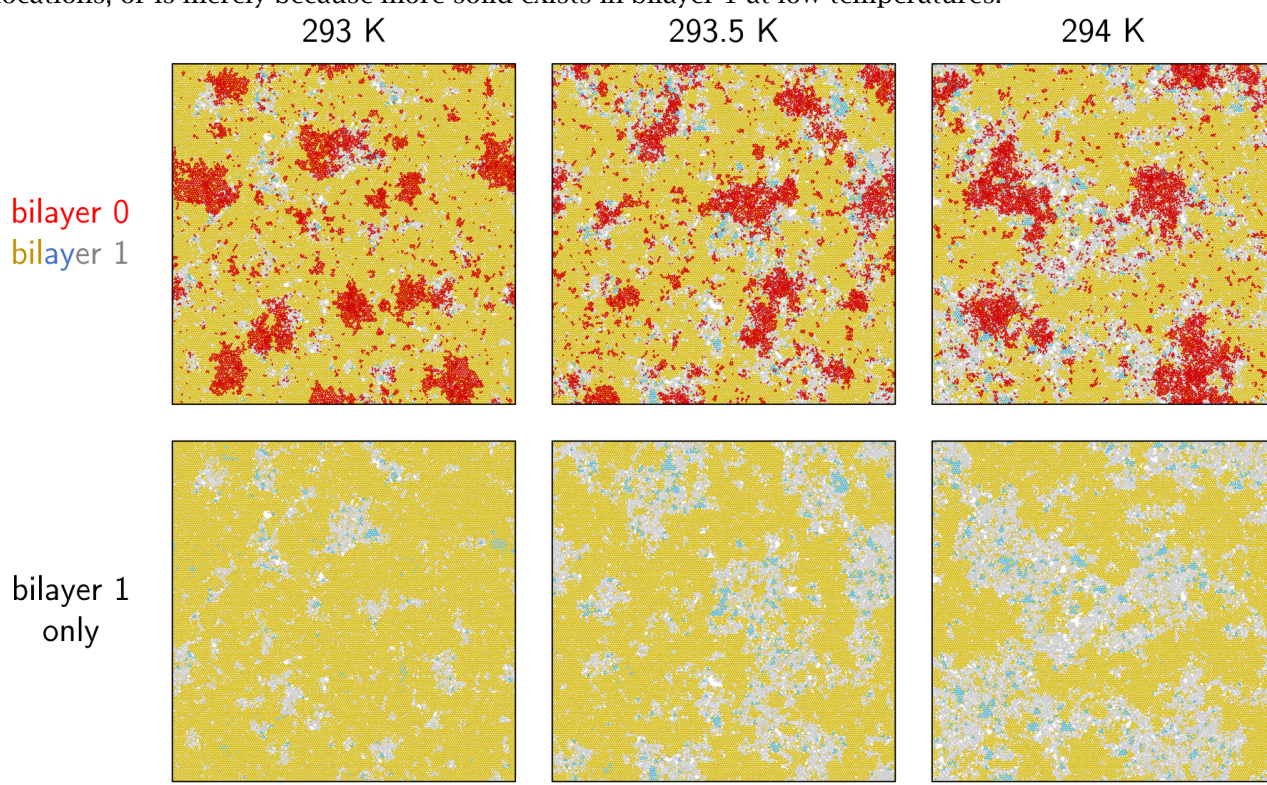

*Figure S10. Ice surfaces with 20% extra molecules (20% means "20% of a full bilayer"; e.g. if a regular system has 10 bilayers and each bilayer has 10,000 molecules, then the current system has 102,000 molecules, with extra 2,000 molecules added on the surface). Upper panels: both bilayer 0 and 1 are shown. Molecules in bilayer 0 (see Supplemental Sec. 6) are colored in red; see legend in Fig. 1 for other molecules. Lower panels: only bilayer 1 is shown. Temperatures are shown at the top.*

*The system size is ~70\*70 nm. Each simulation starts from a perfect ice surface and lasts for 125 ns. Extra molecules are added at the beginning of simulations, by smashing one molecule to the surface every 0.5 ps, with initial velocities of 0.2~0.3 nm/ps towards the surface. The high molecule addition rate is selected in the hope of taking the system away from the initial structure, so it is more likely to reach a thermodynamic stable state later (though it is not guaranteed).*

When more extra molecules exist on the surface, clusters grow further along with the liquid region beneath them (Fig. S11(a1-a2)), and may connect together and isolate the solid regions in bilayer 1 (a1-b2). Following this trend, the clusters will finally occupy the whole bilayer 0 (c1-d2); at this moment it may form either a premelted or solid surface ((d) shows a premelted surface), depending on temperature and kinetics details.

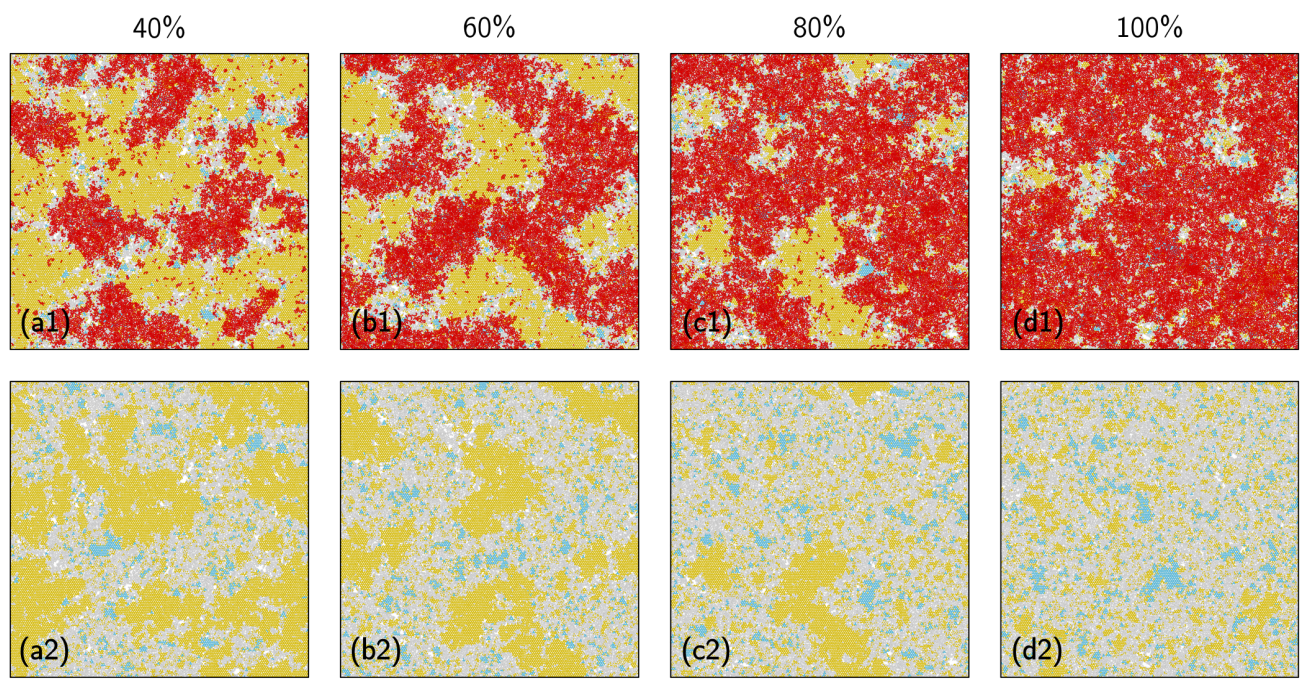

*Figure S11. Ice surface at 293.5 K, with 40%~100% extra molecules (shown at top). They come from four simulations: each run starts from a perfect ice surface and molecules are added separately, but the random seeds for initial velocities and molecule addition are same.*

In the main text, we have seen that the premelting transition is primarily driven by holes. Since excess molecules on the surface fill the holes, we may expect the unpremelted phase becomes more stable. However, detailed studies on this could be difficult, as the coexistence technique becomes unreliable (the "cluster on liquid" phase is inhomogeneous on such a large scale that cannot be neglected at general MD simulation sizes).

# 9 The static premelting profile of (10-10) and (11-20) planes

This work mostly focuses on the basal plane (0001) of ice, which is commonly exposed on ice surfaces. The ice crystal may expose other faces too, namely the primary (10-10) and secondary (11-20) prismatic faces. Fig. S12(a) shows the static premelting behavior of ML-mW ice on the (10-10) plane. Notably, the premelting on the (10-10) plane looks like a continuous transition: there is a narrow range of "transition temperature" around 294 K where the first 2 bilayers lose their order, but the order parameters change continuously in this region. Besides, the 3$^{rd}$ bilayer loses its order at ~295.2 K, close but slightly lower than the melting point. Fig. S12(b) shows the results of the (11-20) plane. There is no "transition temperature" for premelting on this plane: every layer loses its order gradually without significant discontinuity or inflection points. Furthermore, it seems that all layers move toward complete melting "synchronously", indicating a diverging QLL thickness at the melting point. This behavior is significantly different from the basal plane and is observed in some experiments[11-12].

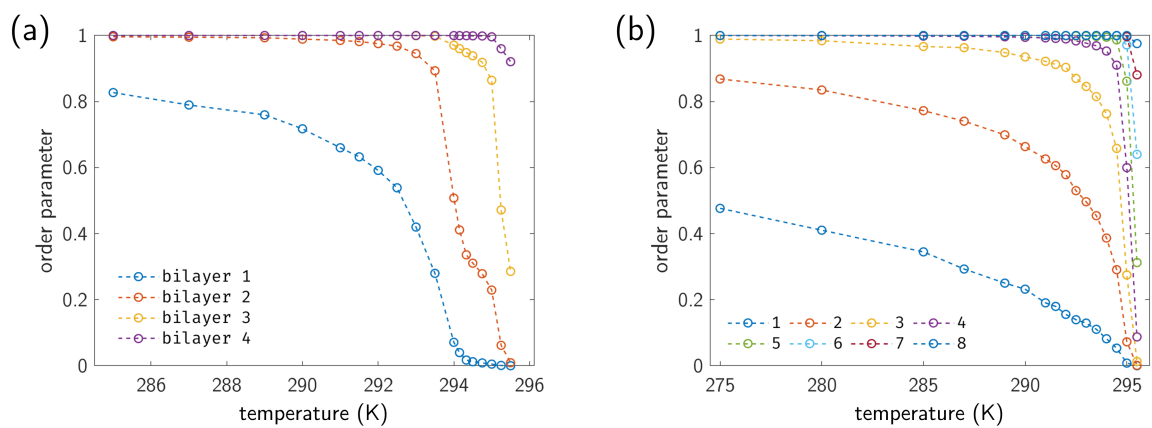

*Figure S12. The static premelting profile for the (a) (10-10) and (b) (11-20) planes. Each line in (b) represents a monolayer of thickness ~0.225 nm (monolayers are equally spaced along the [11-20] direction).*